\documentclass[11pt,preprint]{emulateapj}
\usepackage{psboxit}

\newcommand{\kms}{\,km~s$^{-1}$} 

\newcommand{\Msun}{\mbox{\,$M_{\odot}$}}
\newcommand{\Lsun}{\mbox{\,$L_{\odot}$}}

\def\spose#1{\hbox to 0pt{#1\hss}}
\def\simlt{\mathrel{\spose{\lower 3pt\hbox{$\mathchar"218$}}
     \raise 2.0pt\hbox{$\mathchar"13C$}}}
\def\simgt{\mathrel{\spose{\lower 3pt\hbox{$\mathchar"218$}}
     \raise 2.0pt\hbox{$\mathchar"13E$}}}

\PScommands

\def\simless{\mathbin{\lower 3pt\hbox
	{$\,\rlap{\raise 5pt\hbox{$\char'074$}}\mathchar"7218\,$}}} 
\def\simgreat{\mathbin{\lower 3pt\hbox
	{$\,\rlap{\raise 5pt\hbox{$\char'076$}}\mathchar"7218\,$}}} 

\newcommand{\mrange}{\ensuremath{0.52 - 0.77\Msun}}

\newcounter{thefigs}

\newcounter{thetabs}

\begin{document}

\title{The Stellar Initial Mass Function of Ultra-Faint Dwarf
  Galaxies:\altaffilmark{1}\\ Evidence for IMF Variations with Galactic
  Environment}

\author{Marla~Geha\altaffilmark{2},
Thomas M. Brown\altaffilmark{3}, 
Jason Tumlinson\altaffilmark{3}, 
Jason S. Kalirai\altaffilmark{3}, 
Joshua D. Simon\altaffilmark{8}, 
Evan N. Kirby\altaffilmark{4,5}, 
Don A. VandenBerg\altaffilmark{6}, 
Ricardo R. Mu\~noz\altaffilmark{7}, 
Roberto J. Avila\altaffilmark{3}, 
Puragra Guhathakurta\altaffilmark{9}, 
Henry C. Ferguson\altaffilmark{3} 
}
\altaffiltext{1}{Based on observations made with the NASA/ESA {\it Hubble
Space Telescope}, obtained at STScI, which
is operated by AURA, Inc., under NASA contract NAS 5-26555.}

\altaffiltext{2}{Astronomy Department, Yale University, New Haven, CT
06520, USA; marla.geha@yale.edu}

\altaffiltext{3}{Space Telescope Science Institute, 3700 San Martin Drive,
Baltimore, MD 21218, USA;  
tbrown@stsci.edu, tumlinson@stsci.edu}

\altaffiltext{4}{University of Physics \& Astronomy, University of California Irvine, 4129 Frederick Reines Hall, Irvine, CA 92697, USA}
\altaffiltext{5}{Center for Galaxy Evolution Fellow}

\altaffiltext{6}{Department of Physics and Astronomy, 
University of Victoria, P.O. Box 3055, Victoria, BC, V8W 3P6, Canada}

\altaffiltext{7}{Departamento de Astronom\'ia, Universidad de Chile, 
Casilla 36-D, Santiago, Chile}

\altaffiltext{8}{Observatories of the Carnegie Institution of Washington, 
813 Santa Barbara Street, Pasadena, CA 91101, 
USA}

\altaffiltext{9}{UCO/Lick Observatory and Department of Astronomy and 
Astrophysics, University of California, Santa Cruz, CA 95064, USA}

\begin{abstract}

  We present constraints on the stellar initial mass function (IMF) in
  two ultra-faint dwarf (UFD) galaxies, Hercules and Leo IV, based on
  deep {\it Hubble Space Telescope (HST)} Advanced Camera for Surveys
  (ACS) imaging.  The Hercules and Leo IV galaxies are extremely low
  luminosity ($M_V = -6.2$, $-5.5$), metal-poor
  ($\langle$[Fe/H]$\rangle$= $-2.4$, $-2.5$) systems that have old
  stellar populations ($> 11$\,Gyr).  Because they have long
  relaxation times, we can directly measure the low-mass stellar IMF
  by counting stars below the main-sequence turnoff without correcting
  for dynamical evolution.  Over the stellar mass range probed by our
  data, \mrange, the IMF is best fit by a power-law slope of $\alpha =
  1.2_{-0.5}^{+0.4}$ for Hercules and $\alpha = 1.3\pm0.8$ for Leo
  IV. For Hercules, the IMF slope is more shallow than a Salpeter IMF
  ($\alpha=2.35$) at the 5.8-$\sigma$ level, and a Kroupa IMF
  ($\alpha=2.3$ above 0.5\Msun) sat 5.4-$\sigma$ level.  We
  simultaneously fit for the binary fraction, finding $f_{\rm binary} =
  0.47^{+0.16}_{-0.14}$ for Hercules, and $0.47^{+0.37}_{-0.17}$ for
  Leo IV.  The UFD binary fractions are consistent with that inferred
  for Milky Way stars in the same mass range, despite very different
  metallicities.  In contrast, the IMF slopes in the UFDs are
  shallower than other galactic environments.  In the mass range 0.5
  -- 0.8\,\Msun, we see a trend across the handful of galaxies with
  directly measured IMFs such that the power-law slopes become
  shallower (more bottom-light) with decreasing galactic velocity
  dispersion and metallicity.  This trend is qualitatively consistent
  with results in elliptical galaxies inferred via indirect methods
  and is direct evidence for IMF variations with galactic environment.

\end{abstract}


\section{Introduction} \label{sec_intro}

The stellar initial mass function (IMF) parameterizes the relative
number of stars formed in a single age population as a function of
stellar mass.  The IMF is fundamental to all calculations of star
formation rates and galaxy stellar masses \citep[for reviews
see][]{bastian10a, kroupa13a}.  The classic \citet{salpeter55a} IMF is
a single power law with slope $\alpha=2.35$ ($dN/dm \propto
m^{-\alpha}$, where $N$ is the number of stars of mass $m$).
Salpeter's original IMF was based on stars down to 0.4\Msun, and
modern Milky Way studies indicate a break in the IMF slope below this
mass scale.  \citet{kroupa02a} parameterized the Milky Way IMF as a
broken power law, with $\alpha=2.3$ above 0.5\Msun\ and a shallower
$\alpha=1.3$ slope below this mass.  \citet{chabrier03a,chabrier05a}
parameterized the IMF below 1\Msun\ as a log-normal distribution with a
characteristic mass $m_c \sim$0.2\Msun.

One may expect that the low mass IMF should depend on the physical
properties of the stellar birth cloud such as the gas density,
metallicity, or turbulent velocity \citep[e.g.,;][]{larson05a, bate09a, Myers11a,
marks12a, hopkins12a}, however, IMF observations are
largely invariant within the Milky Way \citep{bochanski10a, covey08a}.
Recent indirect studies suggest that the low mass IMF slope does vary
outside the Milky Way and may be a function of the global galactic
potential: studies of integrated line strengths, kinematics and
gravitational lensing studies of elliptical galaxies appear to favor
IMFs that become increasingly bottom-heavy (with IMF slopes similar to
or steeper than Salpeter) toward higher galaxy masses
\citep{treu10a,vD11a, cappellari12a,dutton12a,conroy12a}.

The ultra-faint dwarf (UFD) galaxies are a recent class of diffuse
Galactic satellites discovered in Sloan Digital Sky Survey (SDSS) data
\citep[e.g.,][]{willman05a, belokurov06a}.  The UFDs are the least
luminous \citep[$-8 < M_V < -1.5$;][]{martin08a,munoz10a} and most
dark matter-dominated \citep{simon07a} galaxies known. Their average
metallicities are less than typical globular clusters
\citep{kirby08a}.  Analysis of {\it HST} photometry implies stellar
ages at least as old as the oldest known globular clusters
\citep{brown12a}.  Thus, the UFDs are a prime environment to test
predicted IMF variations with the temperature, density, or cosmic
epoch of the star-forming environment \citep{tumlinson07a}.

Direct estimates of the IMF based on counting resolved main-sequence
stars are largely limited to the nearby Galactic field and star
clusters \citep{bastian10a}.  Milky Way open clusters provide a
relatively large stellar mass range over which to measure the IMF
\citep[0.08 -- 7 \Msun, e.g.,][]{Moraux04a}, but are more metal-rich
than the UFDs.  At metallicities less than [Fe/H] $< -1$, {\it HST}
studies of Galactic globular clusters probe the IMF down to main
sequence masses between $0.2 - 0.7$\,$M_{\sun}$ \citep{paust10a}.
However, dynamical evolution, such as mass segregation and
evaporation, can significantly change the slope of the mass function
\citep{verperini97a, baumgardt03a}.  Interestingly, the observed MF
slopes correlate contrary to expectation with concentration
\citep{deMarchi10a}.  The observed correlation can be understood as
dynamical evolution combined with either gas expulsion of residual gas
\citep{marks12a}, or related to orbital properties and the degree of
tidal stripping (K\"upper et al.~in prep.).

Dwarf galaxies have metallicities similar to or lower than Galactic
globular clusters, but have relaxation times longer than a Hubble time
and therefore do not require corrections for dynamical
evolution. Dwarf galaxies in which the low mass IMF has been directly
measured are the Milky Way satellites Ursa Minor ($M_V = -9.2$, [Fe/H]
= $-2.0$), Draco ($M_V = -8.6$, [Fe/H] = $-2.0$), and the Small
Magellanic Cloud (SMC; $M_V = -15$; [Fe/H] = $-1.2$).  \citet{wyse02a}
used {\it HST}/WFPC2 data to conclude that the Ursa Minor IMF is
consistent with a power law slope $\alpha = 1.8$ over the mass range
$0.4 - 0.7$\,\Msun.  \citet{grillmair98a} found a power law slope for
the Draco dwarf galaxy between $2.1 < \alpha < 2.3$ for an assumed age
of 12\,Gyr, based on {\it HST}/WFPC2 imaging extending to 0.6\Msun.
\citet{kalirai12} used {\it HST}/ACS data to conclude that the IMF of
the SMC has a power law slope of $\alpha = 1.90_{-0.15}^{+0.10}$ over
the mass range $0.37 - 0.93$\,\Msun.

Using the {\it HST}/ACS we are undertaking a deep imaging survey of
UFDs reaching several magnitudes below the main-sequence turnoff. The
program includes the Milky Way satellites Hercules, Leo IV, Ursa Major
I, Bo\"otes I, Coma Berenices, and Canes Venatici II.  In
\citet{brown12a}, we presented a preliminary analysis of the stellar
populations in the first three galaxies for which data had been taken.
We concluded that all stars in these galaxies were older than 11\,Gyr
and that star formation lasted less than 2 Gyr.

Here we present a companion analysis of the IMFs for Leo IV and
Hercules.  Although Ursa Major~I was included in the age analysis of
\citet{brown12a}, the ACS catalog of this galaxy has significantly
fewer stars, and these stars were observed over a much wider area
(specifically, 9 ACS tiles in Ursa Major I, compared to 1 and 2 tiles
in Leo IV and Hercules, respectively).  While Ursa Major I is closer
than either Leo IV or Hercules, it is subject to significant
contamination from image artifacts and field contamination, and does
not provide a good constraint on the IMF.  We therefore do not include
Ursa Major~I in our analysis.

In \S\,\ref{sec_data} we describe the {\it HST}/ACS data and construction
of the IMFs.  In \S\,\ref{sec_results}, we use these data to constrain
the IMF slope and binary fraction over the stellar mass range \mrange.
In \S\,\ref{sec_impl}, we discuss implications of our IMFs for the
UFDs.  In \S\,\ref{sec_comp},  we compare these results to other direct IMF
measurements and discuss them in a broader cosmological context.

\begin{deluxetable}{lllll}
\tabletypesize{\tiny}
\tablecaption{UFD Galaxy Properties and IMF Results\label{t:prop}}
\tablehead{
\colhead{Row} & \colhead{Quantity} & \colhead{Units} &
\colhead{Hercules} & \colhead{Leo IV} \\ 
}
\startdata
(1) & $\alpha$ (J2000)      &  h$\,$:$\,$m$\,$:$\,$s  &  16:31:05    & 11:32:57 \\
(2) & $\delta$ (J2000)       &  $^\circ\,$:$\,'\,$:$\,''$   & +12:47:18  &$-$00:31:00 \\
(3) & (m-M)$_V$   & mag  & $20.90$              & 21.15 \\
(4) & Distance      & kpc   & 135                  & 156 \\
(5) & E(B-V)          &  mag &  0.08                 & 0.06     \\
(6) & $M_{V}$        & mag  & $-6.2\pm0.4$  & $-5.5\pm0.3$   \\   
(7) & $L_{V}$        & \Lsun & $2.6_{-0.8}^{+1.2}\times10^{4}$    &  $1.4_{-0.3}^{-0.4}\times10^{4}$   \\   
(8) & $r_{\rm eff}$  &   pc   &   230                 & 130  \\
(9) & $\sigma$  &   km s$^{-1}$   &   $5.1\pm0.9 $      & $3.3\pm1.7$  \\
(10) & $\langle$[Fe/H]$\rangle$          &  dex  &  $-2.41$           & $-2.54$ \\
(11) &  $t_{\rm relax}$ & years & $3\times10^{12}$ & $2\times 10^{12}$ \\
\hline
\\[-0.5em]
\multicolumn{5}{l}{IMF Results}\\
(12) & Mass range &  \Msun  & 0.52 - 0.76  &  0.54 - 0.77\\
(13) & $N_{\rm star}$ &             & 2350            & 1054    \\
(14) & $\alpha$    &               & $1.2_{-0.5}^{+0.4}$  & $1.3\pm0.8$  \\
(15) & $m_c$         & \Msun   & $0.4^{+0.9}_{-0.3}$        & $0.4^{+2.1}_{-0.3}$ \\
(16) & $f_{\rm binary}$ &          & $0.48^{+0.20}_{-0.12}$ & $0.47^{+0.37}_{-0.17}$ 
\\[-0.5em]
\enddata
\tablecomments{\tiny (1-2) Right ascension and declination taken from
  \citet{martin08a}.  (3-5) The distance modulus, distance and reddening are
  best fitting values from the {\it HST} CMDs.  (5-7) Absolute
  magnitudes and effective radii from \citet{sand09a} and
  \citet{sand10a} for Hercules and Leo IV, respectively.  (7-8)
  Velocity dispersion and average metallicity from \citet{simon07a} and
  \citet{kirby11a}.  (9) The two-body relaxation is calculated using
  the quantities in (5-7) and Equation 1.38 in \citet{bt08}.  (10)
  Stellar mass range (10) and number of stars (11) included in our IMF
  analysis.  (11-13) Estimated IMF parameters in the mass range
  \mrange.  Binary fractions (14) are given for the power law fits.}
\end{deluxetable}

\begin{figure*}
\epsscale{1.05}
\plotone{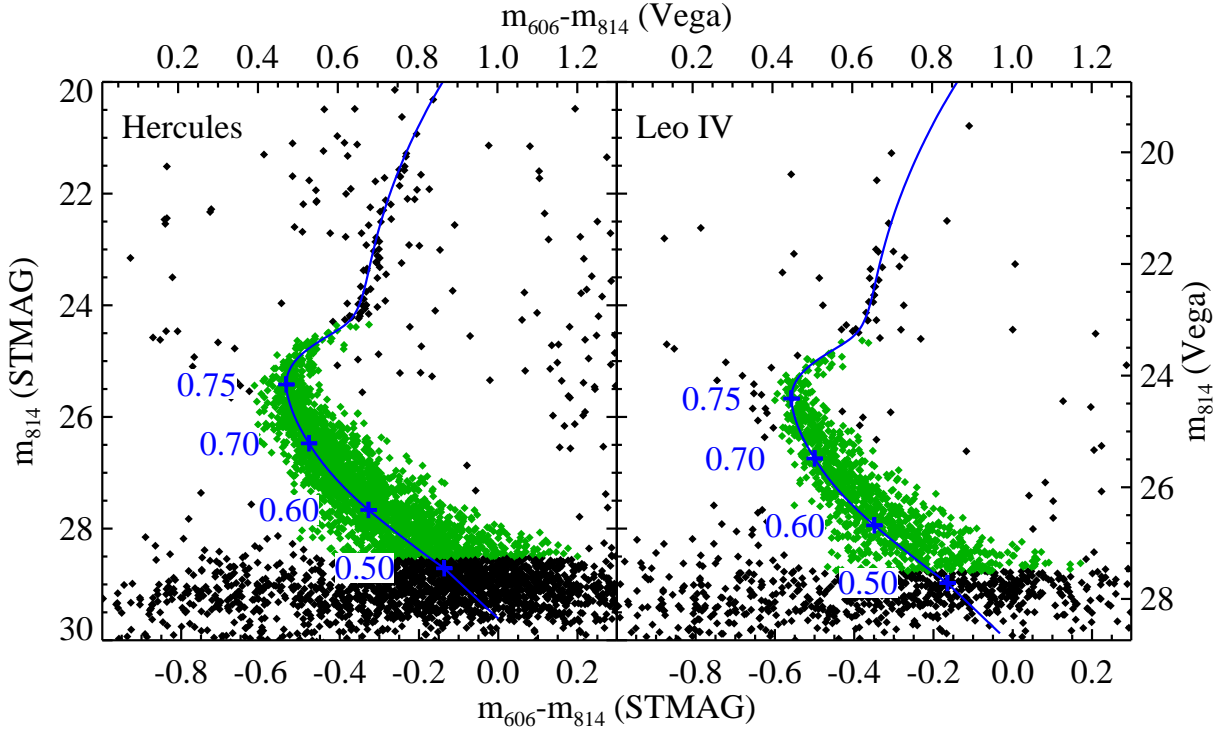}
\caption{The {\it HST}/ACS CMDs of two UFD galaxies, Hercules ({\it
    left}) and Leo IV ({\it right}).  For the IMF analysis, we include
  stars below the sub-giant branch and above the 66\% and 75\%
  completeness limits for Hercules and Leo IV, respectively (green
  points).  The axes are labeled in both STMAG and Vega magnitudes.
  The blue line is a representative isochrone of 13.6\,Gyr and the
  mean metallicity of each galaxy (Table~1).  Blue crosses indicate
  stellar mass in units of \Msun\ on the main
  sequence.  \label{fig_cmds}}
\end{figure*}

\section{Observations and Data Reduction}\label{sec_data}

The Hercules and Leo IV UFDs were discovered by \citet{belokurov06a}
as statistically significant overdensities of stars in the SDSS.  As
listed in Table~1, Hercules and Leo IV have average metallicities of
$\langle$[Fe/H]$\rangle$ = $-2.4$ and $-2.5$, respectively, and show
internal metallicity spreads of more than 0.5\,dex \citep{kirby08a,
  aden09a}.  Both galaxies contain a population of RR Lyrae stars,
implying the presence of stars that are at least as old as 10\,Gyr
\citep{moretti09a,musella12a}.  Ground-based imaging further suggested
old stellar populations
\citep{coleman07,sand10a,dejong10a,Okamoto12a}.  \citet{coleman07} and
\citet{deason12a} suggest that Hercules is tidally disrupting due to
its elongated shape and velocity gradient at large radius.  Since
tidal processes are independent of stellar mass this should not affect
the IMF analysis, and our {\it HST} observations described below are
well within the gravitationally bound region of the object.  Based on
the same {\it HST} observations described below, \citet{brown12a}
confirmed that stars in these two UFDs are exclusively 11\,Gyr or
older.

We obtained deep optical images for Hercules and Leo IV using the
F606W and F814W filters on the {\it HST}/ACS Wide Field Camera between
August 2011 and January 2012 (GO-12549, PI:\,Brown).  The total
exposure times were 25625 and 41060~seconds, respectively.  We
obtained two ACS tiles for Hercules and one ACS tile for Leo IV.  All
images were dithered to mitigate detector artifacts and enable
resampling of the point spread function (PSF).  Our image processing
includes the latest pixel-based correction for charge-transfer
inefficiency \citep{anderson10a}.  We co-added the images for each
filter in a given tile using the IRAF DRIZZLE package
\citep{fruchter02a}, with masks for cosmic rays and hot pixels derived
from each image stack, resulting in geometrically-correct images with
a plate scale of 0.03$^{\prime\prime}$ pixel$^{-1}$ and an area of
approximately $210^{\prime\prime} \times 220^{\prime\prime}$.

We performed both aperture and PSF-fitting photometry using the
DAOPHOT-II package \citep{stetson97a}, assuming a spatially-variable
PSF constructed from isolated stars. The final catalog combined
aperture photometry for bright stars with photometric errors $<0.02$
mag and PSF-fitting photometry for the rest, all normalized to an
infinite aperture. Due to the scarcity of bright stars, the
uncertainty in the normalization is $∼0.02$ mag.  Our photometry is in
the STMAG system: $m_\lambda= -2.5 \times $~log$_{10} f_\lambda -21.1$, except
where we explicitly state Vega magnitudes (Figure~\ref{fig_cmds}).
The catalogs were cleaned of stars with poor photometry and background
galaxies. Sources were rejected based on photometric errors ($<
0.1$\,mag) and the DAOPHOT \mbox{$\chi$} and $sharp$ parameters.  We
also reject stars with bright neighbors and those falling within the
profiles of extended background galaxies because these are noisier
than isolated stars of comparable magnitude.  We apply the same
rejection criteria in determining the completeness of our data
described below.  For more details on the data reduction see
\citet{brown09a}.

We performed extensive artificial star tests to evaluate the
photometric scatter and completeness for each galaxy.  These tests
employed the same PSF model and PSF-fitting routines used in the
construction of the observed catalogs, including the algorithms for
culling poor-quality stars, image artifacts and corrections for charge
transfer efficiency.  Artificial stars were inserted into the image
with appropriate reductions in signal (and noise) due to charge
transfer inefficiency.  Stars were then blindly recovered and
corrected for charge transfer \citep{anderson10a}.  The artificial
stars were inserted over a wide range of color ($-1.3 \le m_{606} -
m_{814} \le 0.5$) and magnitude ($32 \ge m_{814} \ge 16$) with most of
the stars falling near the observed stellar locus and biased toward
fainter magnitudes, thus providing the most fidelity in the analysis
region.  A total of 5,000,000 artificial stars were inserted into each
image, spread over thousands of passes in order to avoid significantly
altering the level of crowding and the associated photometric scatter.
The number of stars recovered from all passes sets the completeness
fractions given in Table~2.  For this analysis, we include stars
fainter than the red giant branch and brighter than the magnitude
where photometric errors approach the main-sequence width
and photometric artifacts begin to contribute significantly to the
catalog (green region in Figure~\ref{fig_cmds}, see Table~2).  The
faint cutoff corresponds to the 66\% and 75\% completeness limits for
Hercules and Leo IV, respectively.  This magnitude region corresponds
to a stellar mass range of \mrange\ and includes 2380 stars in Hercules
and 1054 stars in Leo IV.

\begin{figure*} 
\epsscale{1.05} 
\plotone{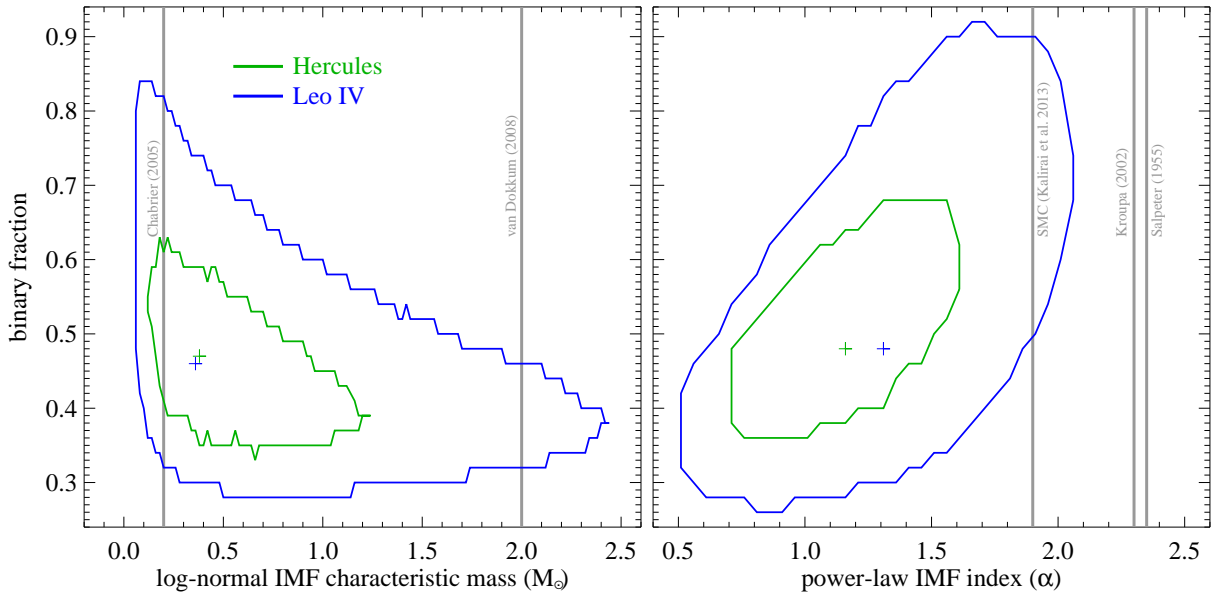}
\caption{One-sigma confidence contours for the Hercules (green) and
  Leo IV (blue) UFD galaxies.  For the mass range probed by our data
  (\mrange), we plot our results for a log-normal IMF ({\it left}) and
  power-law IMF ({\it right}). The best fitting combination of
  characteristic mass ($m_c$) or slope ($\alpha$) and stellar binary
  fraction ($f_{\rm binary}$) are shown as plus symbols.  For
  reference, we indicate values of $m_c$ and $\alpha$ for IMFs in the
  literature.  \label{fig_imf}}
\end{figure*}

\subsection{Constructing the Model IMFs}\label{ssec_construct}

We analyze our observed luminosity functions by forward modeling
stellar evolutionary tracks with an analytic form of the IMF and our
observational errors.  The free parameters in our models are the IMF
shape and the binary fraction.  We also allow the distance and
reddening of each UFD galaxy to float within the one-sigma errors of
the values determined in \citet{brown12a}.  Because our IMF fits
exclude stars brighter than the subgiant branch, the resulting values
of $(m-M)_V$ and E(B-V) are slightly different than \citet{brown12a}
at the hundredth of a magnitude level, providing a better fit to the
main-sequence region.  These values are listed in Table~1.  We fix the
galactic age to 13.6\,Gyr, as determined in \citet{brown12a}.  We
further fix the distribution of metallicities in each UFD galaxy based
on the spectroscopic metallicity distribution functions determined by
\citet{kirby08a} and \citet{kirby11a}.  We explore these assumptions
further below.

We simulate the observations beginning with new Victoria-Regina
evolutionary tracks (VandenBerg et al., in prep.)~which assume a
solar heavy-element mixture by \citet{asplund09}, with 0.4\,dex
enhancements in the $\alpha$-element abundances, and then scaled to
the [Fe/H] values of interest.  These are computed using the same code
 described in \citet{vandenBerg12a}.  These isochrones are
transformed into the observed STMAG magnitude system using the MARCS
model atmosphere library and the throughput curves for the ACS F606W
and F814W filters. The transformation is calibrated to observations of
globular clusters in the same filters, with 1\% agreement over the
main sequence, subgiant branch, and RGB.

We assume two different functional forms of the IMF: a single
power-law with free parameter $\alpha$ (where a Salpeter IMF is
$\alpha = 2.35$, see Introduction for the functional form), and a
log-normal model with parameters $\sigma$ and characteristic mass
$m_c$ (where a Chabrier IMF for single stars is $\sigma = 0.69$ and
$m_c = 0.2$\Msun). In the case of the log-normal IMF, our stellar mass
range is too small to determine the distribution width and, for the
purpose of this paper, we keep this value fixed at the Chabrier value
of $\sigma = 0.69$.  We note that for the Milky Way IMF, current data
cannot significantly differentiate between these two function forms
\citep{kroupa13a}.

The shape of the IMF is somewhat degenerate with the presence of
unresolved binary stars \citep{kroupa91a}.  We have therefore opted to
fit both the IMF shape and the binary fraction simultaneously. We
define the binary fraction, $f_{\rm binary}$, as the fraction of
unresolved systems which are binary (e.g., if 25 out of 100 systems
are binary, then $f_{\rm binary} = 0.25$ and the total number of
individual stars is 125). We draw binary companions from the full
distribution of stellar masses, meaning that the mass ratio
distribution is flat.  Because we are explicitly fitting for binary
fractions, the reported IMF parameters below pertain to the
single-star IMF, rather than the system IMF.

We convolve the IMF with our known observational errors and
completeness estimates, resulting in a stellar probability
distribution for a single age and metallicity.  While the UFDs likely
had a period of star formation which lasted at least 100\,Myr and at
most 2\,Gyr \citep{vargas13a, brown12a}, our assumption of a single
age has insignificant effects on the IMF, because nearly all of the
weight in our fit comes from stars below the main-sequence turnoff of
the oldest population which change by less than 0.01\,mag over this
time period.  For the fixed single age of each UFD, we
linearly combine tracks for different metallicities to recreate the
spectroscopic metallicity distribution function (MDF) as determined by
\citet{kirby11a} based on Keck/DEIMOS data from \citet{simon07a}.  The
UFDs have significant internal metallicity dispersion, over 0.6 and
0.7\,dex for Hercules and Leo IV, respectively.  Since the MDFs used
in this analysis are based on fewer than 25 stars in each galaxy, we
roughly estimate the IMF error which may be introduced by metallicity
errors.  We simulate two CMDs based on our observational errors and
completeness values, assuming a Salpeter IMF.  For one simulated
dataset, we use the observed Hercules MDF.  For the second dataset, we
assume a single metallicity of [Fe/H]$=-2.4$.  We then recover the
IMFs assuming the same single metallicity isochrone.  The IMF slopes
for these two simulations are less than 0.1 different in the power law
slope which is far less than the 1-$\sigma$ error on this parameter
determined below.
 
\begin{figure*} 
\epsscale{1.0} 
\plotone{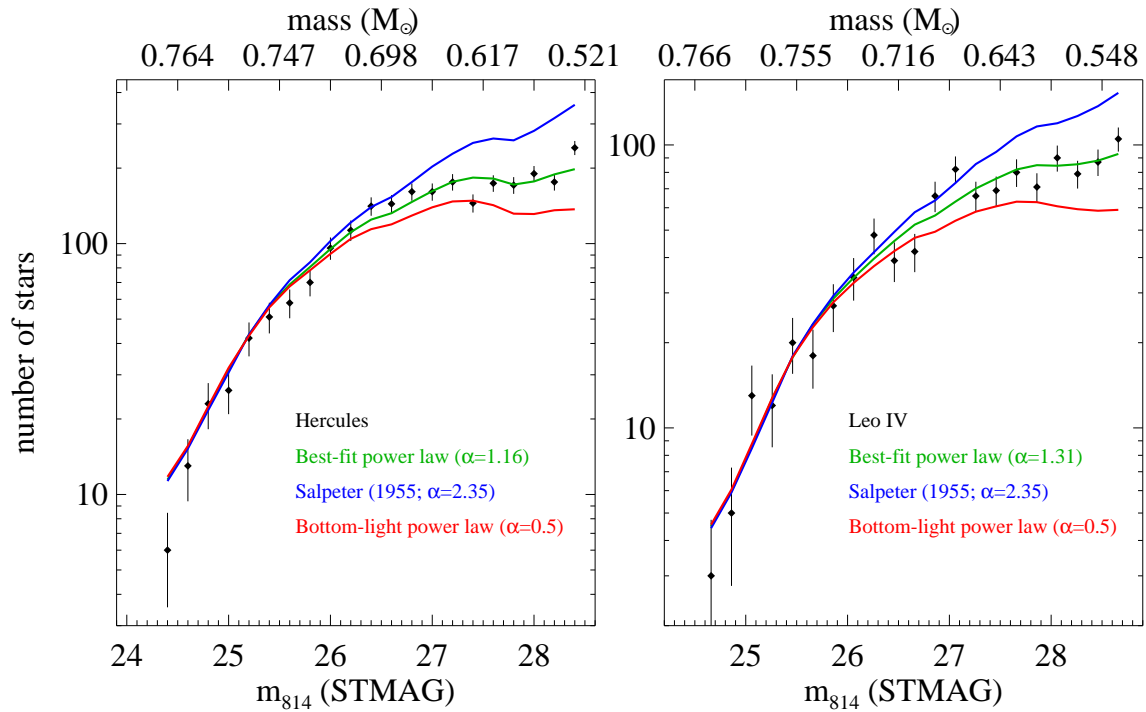}
\caption{The observed luminosity function for Hercules ({\it left})
  and Leo IV ({\it right}).  Errors bars are computed from the
  observed number of stars in each luminosity bin.  For comparison, we
  plot three theoretical power law IMFs, convolved with our
  observational errors and photometric completeness. The fits were
  normalized to reproduce the number of stars in the observed
  luminosity function, but here they have been normalized at the bright end for
  clarity.  We compare our best-fitting model (green) to a Salpeter
  IMF ($\alpha=2.35$, blue) and an extremely bottom-light IMF
  ($\alpha = 0.5$, red).  \label{fig_imf2}}
\end{figure*}

The final comparison between model and observations is done in CMD
space by binning each into Hess diagrams.  The Hess diagrams have axes
of magnitude and color with 0.1\,mag bins in each axis.  The model and
observed Hess diagrams are then compared using the Maximum Likelihood
statistic of \citet{Dolphin02a}, using the stars defined as members of
each galaxy (Figure~\ref{fig_cmds}).  We search a grid over two free
parameters: the IMF slope ($\alpha$ for a power-law IMF) or
characteristic mass ($m_c$ for a log-normal IMF) and the binary
fraction $f_{\rm binary}$.  The best-fit model is that with the
minimum Maximum Likelihood statistic over the two free parameters
(Figure~\ref{fig_imf2}).

We do not correct our inferred IMFs for dynamical evolution.
Dynamical two-body relaxation processes, such as mass segregation and
evaporation, are mass dependent and will alter the initial stellar
mass function over a relaxation timescale, which we calculate
according to Equation 1.38 in \citet{bt08}: $t_{\rm relax} \sim
(N/log_{10}N) * (R/\sigma)$.  We assume the values for the radius ($R$) and
stellar velocity dispersion ($\sigma$) using the values listed in Table ~1.  The
resulting timescales are on the order of $10^{12}$ years, a few
hundred times the age of the Universe.  Thus, two-body relaxation
processes can safely be ignored in the UFD galaxies.

\section{Results}\label{sec_results}

The {\it HST}/ACS imaging for Hercules and Leo IV extends roughly three
magnitudes below the main-sequence turnoff (Figure~\ref{fig_cmds}).
Stars included in the IMF analysis correspond to stellar masses
between \mrange.  While this is not a particularly large range in
stellar mass, it is sufficient to distinguish between various IMFs
suggested in the literature.

\subsection{Constraints on the IMF}\label{ssec_imf}

In Figure~\ref{fig_imf}, we plot the one-sigma error contours for the
binary fraction and IMF parameter $\alpha$ or $m_c$, in the case of
the power-law and log-normal IMF, respectively.  The constraints for
Hercules (green) are tighter as compared to Leo IV (blue) since
Hercules is intrinsically more luminous and the ACS photometric
catalog is better populated along the main sequence.  The best-fit solutions
are consistent between the two galaxies.  Based on the maximum
likelihood statistic, the power law and log-normal models provide an
equally good fit to the data: the log-normal form is slightly
preferred but at low (0.3$\sigma$) significance.

\begin{figure} 
\epsscale{1.25} 
\plotone{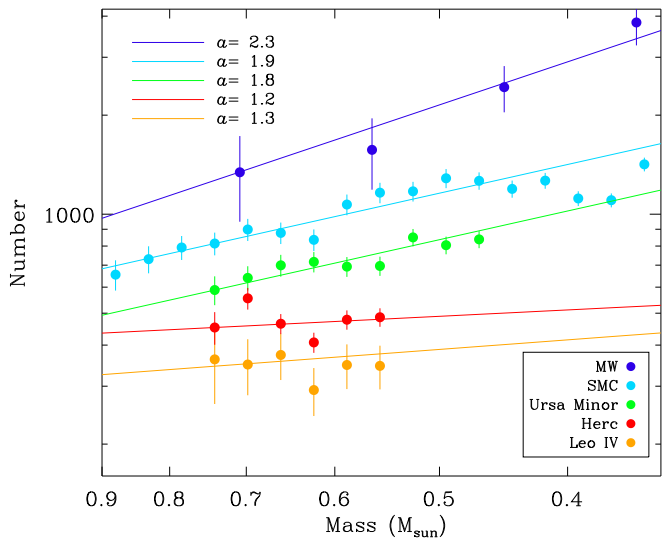}
\caption{Stellar mass functions for the five galaxies in which the IMF
  has been measured via direct star counts: the Milky Way (blue,
  Bochanski et al.\,2010), the SMC (light blue, Kalirai et al.\,2012),
  Ursa Minor dSph (green; Wyse et al.\,2002), Leo IV (orange; this
  work) and Hercules (red; this work).  Except for Hercules, the
  vertical normalization is arbitrary.  For reference, the published
  power law slopes are shown for each dataset, normalized at 0.75\Msun.
  We note that a power law slope of $\alpha = 1$ is a flat line in
  this log-log plot.  The UFD galaxies show noticeably flatter mass
  functions in this mass range.  \label{fig_mass}}
\end{figure}

For a single power law model, we find a best fit slope of $\alpha =
1.2_{-0.5}^{+0.4}$ for Hercules and $\alpha = 1.3\pm0.8$ for Leo IV
(Figure~\ref{fig_imf}, right panel).  In Figure~\ref{fig_imf2}, we
plot the observed luminosity functions for each UFD and overplot our
best-fitting model (green) compared to a Salpeter IMF ($\alpha=2.35$,
blue) and an extremely bottom-light IMF ($\alpha = 0.5$, red).  For
Hercules, the IMF slope is shallower than the Salpeter ($\alpha=2.35$)
and Kroupa ($\alpha=2.3$ above 0.5\Msun) IMF at the 5.8-$\sigma$ and
5.4-$\sigma$ levels.  For Leo IV, the Salpeter and Kroupa values are
1.9-$\sigma$ and 1.7$\sigma$ from our best fit value.  We note that
our error contours are non-Gaussian.  While we strongly rule out the
Salpeter and Kroupa values for Hercules, we note that just below our
mass range of \mrange\, the Milky Way IMF slope turns over.  Below
0.5\Msun, the Kroupa IMF slope changes to $\alpha=1.3$, which is well
within the UFD values.  The exact location of this turn over in the
Milky Way is unclear.  For example the data from \citet{bochanski10a}
plotted in Figure~\ref{fig_mass} suggests a Salpeter-like slope down
to 0.3\Msun.  We will discuss this further in \S\,\ref{sec_comp}.  Our
power-law slopes are also shallower than recent IMF results in the
Small Magellanic Cloud. \citet{kalirai12} found $\alpha =
1.90_{-0.15}^{+0.10}$ over the mass range $m = 0.37 - 0.93$\,\Msun,
which is 2.3-$\sigma$ away from the Hercules IMF.  These authors also
fit a power law only to stars more massive than 0.6\Msun, finding a
slope $\alpha = 2.1\pm 0.3$.

We alternatively fit a log-normal function to our UFD data, showing
the one-sigma confidence intervals in the left panel of Figure~\ref{fig_imf}.  The
best-fitting log-normal IMF values are: $m_c =
0.4^{+0.9}_{-0.3}$\Msun\ for Hercules and $m_c =
0.4^{+2.1}_{-0.3}$\Msun\ for Leo IV.  The Milky Way value of $m_c =
0.2$\,\Msun\ \citep{chabrier05a} is within our 1-$\sigma$ errors.
However, the fact that the data prefer a more massive characteristic
mass is consistent with the results for our power law fits, since, in
our mass range, a more massive characteristic mass provides an overall
flatter IMF shape as determined for the power law slope.  In addition,
the data for Hercules rule out an extremely 'bottom-light' IMF, with
$m_c \sim 2$\Msun, as favored by \citet{vD08a} for massive elliptical
galaxies, although this result has been subsequently disputed
\citep{vD11a}.

In Figure~\ref{fig_mass}, we plot the mass function of our UFDs.  We
compute the mass function by converting the completeness corrected
F814W luminosity function (Table~2) into stellar mass assuming the
best fitting single Victoria-Regina isochrone and rebinning
using constant logarithmic mass bins.  While this method does not
fully take into account metallicity spread or photometric errors
(which are accounted for in our formal analysis,
\S\,\ref{ssec_construct}), this effect is small and allows us to
compare more directly to literature data.  Published luminosity
functions are available for two other galaxies with low mass IMF analyses.
We convert the luminosity functions for the SMC (Table~1 in Kalirai et
al.~2012) and the dwarf spheroidal galaxy Ursa Minor (Table~9 in Wyse
et al.~2002) into mass functions in the same manner as the UFDs based
on the best-fitting single isochrone.  Literature data for the mass
function itself is available only for the Milky Way and we plot data from
Table~11 in \citet{bochanski10a} based on SDSS photometry of Milky Way
field stars.  We have arbitrarily renormalized the mass functions and
overplot representative analytic fits.  While Figure~\ref{fig_mass}
emphasizes the limited stellar range over which the UFD IMFs have been
measured, it visually confirms that the UFD IMF slopes are shallower
than the observed Milky Way and SMC IMFs.  We explore the implications
of a shallow IMF for the UFDs in \S\,\ref{sec_impl} and discuss these
results in the context of other galaxies in \S\,\ref{sec_comp}

\begin{figure*}
\epsscale{1.15}
\plotone{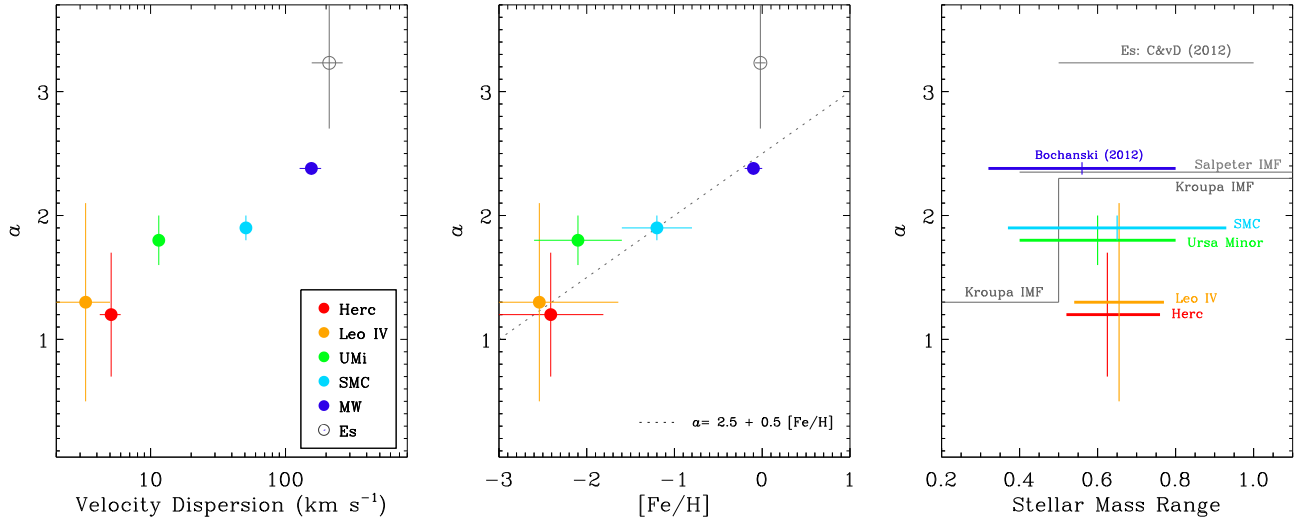}
\caption{The power law slope, $\alpha$, plotted against galaxy
  velocity dispersion  ({\it left}), metallicity [Fe/H] ({\it middle}) and the observed stellar mass range ({\it right}).  Data are taken from the same sources as
  Figure~\ref{fig_mass}.  The dotted line in the middle panel is an
  empirical relationship suggested by \citet{kroupa01a} with a
  zero-point shift to fit these data.   Indirect measurements are shown for
  elliptical galaxies from \citet{conroy12a}.  A trend
  is seen in the sense of shallower, more bottom-light IMF slopes
  towards less massive and more metal-poor galaxies.
  \label{fig_lit}}
\end{figure*}

\subsection{Constraints on the Binary Fractions}

The presence of unresolved binary stars can mimic a flattening of the
IMF at low masses \citep{kroupa91a, bochanski10a}.  On the main
sequence, \citet{kroupa91a} first demonstrated that unresolved binary
systems widen the main sequence beyond observational errors, with
binary systems brighter and redward of the single star main sequence.
Our best-fitting single-power law IMF is degenerate between steeper
IMF slopes with high binary fractions and shallower (more
bottom-light) IMF slopes with lower binary fractions
(Figure~\ref{fig_imf}). For Hercules, the binary fraction is
constrained to be $f_{\rm binary} = 0. 48^{+0.20}_{-0.12}$ for a
single power law or $0.47^{+0.16}_{-0.14}$ for a log-normal IMF. The
binary fractions for Leo IV are similar, but far less constrained due
to the smaller number of observed stars (Table~1).  While binary stars
were known to exist in the UFDs based on repeated spectroscopic
measurements of red giant branch stars \citep{simon11a, koposov11a},
these observations did not strongly constrain the binary fraction
itself \citep{martinez11a}.  Our results are the first quantitative
constraints on the binary fraction in the UFDs.

The binary fractions inferred for the UFDs pertain to stars in the
mass range \mrange, predominantly K-dwarf stars.  In the solar
neighborhood, the multiplicity of stars decreases with decreasing
stellar mass \citep{kraus12a}, with more massive OB-type stars having
binary fractions upwards of $f_{\rm binary} = 0.7$ \citep{peter12a},
down to M-stars with binary fractions between 0.2-0.3
\citep{janson12a}.  This trend may be the result of intrinsic
differences in the binary fraction as a function of stellar mass, the
consequence of a single overall binary fraction and random pairings
across the mass spectrum, or possibly due to binary disruption over
time combined with ages differences between spectral populations
\citep{kroupa93a,marks11a}.  For K-dwarfs, \citet{duquennoy91a}
suggest that roughly half of solar neighborhood K-dwarf systems are
binary; a recent analysis by \citet{raghavan10a}~find a solar
neighborhood binary fraction for FGK stars of $0.46\pm0.02$.  This is
in remarkable agreement with our UFD binary fractions, despite the
significantly lower metallicity environment of the UFDs as compared to
the solar neighborhood.  While noting that binary fractions as high as
$f_{\rm binary} = 0.68$ are allowed within our 1$-\sigma$ limits for
Hercules, the best-fitting binary fractions appear to be a more
'universal' property than the slope of the IMF \citep{marks11a}.

\section{Implications of a Shallow IMF in the UFDs}\label{sec_impl}

The two UFDs presented have shallower IMF slopes in the stellar mass
range \mrange\ than expected from a Salpeter or Kroupa IMF and are
marginally shallower than a Chabrier IMF over the observed mass range.
This difference is seen visually in Figure~\ref{fig_imf2}.  For an IMF
slope $\alpha = 2.3$, we would have expected to observe 2900 stars as
compared to the 2380 stars observed for Hercules with an inferred IMF
slope of $\alpha = 1.2$.  Many properties of the UFDs have been
calculated assuming a standard IMF.  Our results have implications for
the stellar mass, mass-to-light ratio supernova rates and formation
models of these low luminosity systems.

\citet{martin08a} calculated the stellar mass of Hercules and Leo IV
assuming both a Kroupa and Salpeter IMF.  Since the Kroupa IMF has a
shallower IMF slope below 0.5\Msun, the resulting total stellar masses
are 50\% that of a Salpeter IMF.  If we instead use our IMF to
determine stellar mass, assuming that a power law slope of $\alpha =
1.3$ applies over the full mass spectrum, our calculated stellar
masses have 40\% the mass of a Salpeter IMF.   Alternatively, the
best-fit log-normal fit with $m_c=0.4$\Msun\ has 44\% the stellar mass
as a Chabrier IMF with $m_c=0.2$\Msun.   

Shallower IMF slopes have implications for models of UFD galaxy
formation.  An IMF is assumed in calculating the amount of available
supernova energy which is an important physical process in many galaxy
formation models \citep[e.g.,][]{governato12a,wyithe13a, teyssier13a},
particularly for dwarf galaxies.  The magnitude of this effect depends
in large part on the behavior of the IMF outside our observation
window.  If the shallow slope measured at low masses applies across
the whole mass spectrum, the number of supernovae expected per
luminous star would {\it increase} over a Salpeter or Kroupa IMF.
Assuming a Milky Way-like IMF for the UFDs under-estimates the effects
of supernova feedback.

\section{Comparisons to Other IMF Studies}\label{sec_comp}

The IMF has been measured via direct star counts in the stellar mass
range 0.5 to 0.8\Msun\ for five distinct galaxies: the Milky Way, SMC,
Ursa Minor, and the two UFDs presented here.  As discussed in
\S\,\ref{ssec_imf}, we directly compare the observed mass functions
for these systems in Figure~\ref{fig_mass}.  We next compare the
published analytic fits to the IMF, focusing for simplicity only on
the power law slope ($\alpha$).

In Figure~\ref{fig_lit}, we plot the power-law IMF slope as a function
of the galactic velocity dispersion ($\sigma$), average metallicity
([Fe/H]) and stellar mass range over which each measurement was made.
The velocity dispersion and metallicity of our UFDs are given in
Table~1.  The SMC IMF results are taken from \citet{kalirai12}, with
the galactic velocity dispersion and metallicity from
\citet{harris06a}.  IMF results for the Ursa Minor dwarf galaxy are
from \citet{wyse02a}, with the galactic velocity dispersion from
\citet{wolf10a} and metallicity from \citet{kirby11a}.  While
\citet{wyse02a} do not provide an error bar on the IMF slope, we
calculate this value by fitting a linear function to the data shown in
Figure~\ref{fig_mass}.  For the Milky Way, we use the fitted IMF
values from \citet{bochanski10a} based on a sample of Milky Way field
dwarfs spanning [Fe/H] = $-0.1$ to $-0.6$.  We plot the Milky Way at a
velocity dispersion of 220/$\sqrt{2}$ \kms, assuming the disk is
embedded in an isothermal halo \citep{burstein97a}.

Figure~\ref{fig_lit} shows a clear trend in the IMF power law slope as a
function of galactic velocity dispersion and metallicity.  The power
law slope becomes increasingly shallow (bottom-light) and less
Salpeter-like with decreasing galaxy mass/metallicity.  This confirms
the visual impression observed in Figure~\ref{fig_mass}.  As noted in
\S\,\ref{ssec_imf},  Hercules is 5.4-$\sigma$ different than the Kroupa
IMF and 2.3-$\sigma$ away from the SMC value.  This is the first clear
evidence for IMF variations with galactic environments based on
direct star counts.

We have not included globular clusters in our comparison in
Figure~\ref{fig_lit}.  The present-day mass functions of many globular
clusters have been measured down to 0.3-0.4\Msun\, and these systems
are of comparable metallicity to the UFD galaxies.  \citet{paust10a}
and others have measured the present day IMF of Milky Way globular
clusters, suggesting that globular clusters begin with a Milky
Way-like IMF and that the observed variations are related to dynamical
evolution.  However, this does not fully explain the observed
correlations and additional mechanisms, such as gas-expulsion from
initially compact clusters \citep{marks12a} or orbital properties
(K\"upper et al.~2013), are needed.  Because stars clusters do not
form within their own dark matter halo, it is possible that star
formation of these systems is more influenced by their parent galaxy.
For these reasons, we do not include star clusters in our comparisons
with the UFDs.

The trends seen in Figure~\ref{fig_lit} are qualitatively consistent
with results in elliptical galaxies inferred via indirect methods
(i.e., based on integrated galaxy light).  Correlation of IMF slope
with galactic velocity dispersion has been shown via spectroscopic
line widths \citep{vD11a,conroy12a, ferreras13a}, fundamental plane
relations \citep{cappellari12a, dutton12a} and gravitational lensing
\citep{treu10a}.  These studies suggest that the most massive
ellipticals have bottom-heavy IMFs, with IMF slopes steeper than
Salpeter.  For example, \citet{cappellari12b} find a systematic trend for a sample
of 260 ellipiticals with velocity dispersions between 65 - 250\kms\ such
that the inferred IMF is closer to Kroupa/Chabrier for low velocity
dispersion ellipticals, and Salpeter or steeper at highest
dispersions.  In Figure~\ref{fig_lit}, we include for comparison the
inferred IMF slopes from Conroy \& van Dokkum (2012,
priv.~communication) based on 35 elliptical galaxies.  These authors
model the IMF as a broken power-law.  We plot their `$\alpha$2', the
IMF slope between 0.5 - 1.0\Msun, a comparable range as our UFD
observations.  The plotted error bars are the sample standard
deviation.

A critical question raised by Figure~\ref{fig_lit} is which physical
property is responsible for the observed trends with IMF slope.  While
we see a clear trend of increasing (steeper) IMF slopes with both
increasing galactic velocity dispersion and increasing metallicity,
these two physical properties are correlated \citep{tremonti04a}.
Metallicity has often been cited as an expected driver of IMF
variation, such that lower metallicity gas clouds fragment into fewer
low mass protostars, and thus have a more shallow IMF.
\citet{kroupa01a} first noted a possible correlation between the IMF
slope and metallicity for Milky Way clusters and suggested an
empirical relation of $\alpha = 2.3 + 0.5\,$[Fe/H].  The slope of this
relationship fits the trend seen in the middle panel of
Figure~\ref{fig_lit} and we have adjusted the zero-point by 0.2\,dex
to better fit the data.  For massive ellipticals, the IMF slope appears
to correlate more strongly with the alpha-element abundance [Mg/Fe]
than either [Fe/H] or galactic velocity dispersion \citep{conroy12a}.
Since alpha-abundances are related to supernova Type II enrichment, it
is plausible that star formation rates or specific star formation
density is the controlling physical parameter \citep{weidner05a}.  For
the UFD galaxies which have $\sim10^4$\Lsun\ and formed stars over a
period between 100\,Myr to 2\,Gyr \citep{brown12a, vargas13a}, the
star formation rates (SFR) are between $10^{-5} < {\rm SFR} <
10^{-4}$\Msun\,year$^{-1}$.  The Milky Way has a SFR of roughly
1\,\Msun\,year$^{-1}$ \citep{robitaille10a} and the most massive
galaxies with the steepest inferred IMFs by \citet{conroy12a} have
inferred SFRs up to 100\,\Msun\,year$^{-1}$. Determining the physical
properties which control the IMF in different galaxies is a key
question for future studies.

\section{Conclusions}\label{sec_concl}

We have directly measured the IMF in two UFDs via {\it HST}/ACS star
counts.  Since the UFDs are an ancient, metal-poor and nearly single age
population \citep{brown12a} with long two-body relaxation times, these
objects provide a unique environment in which to determine the
low-mass stellar IMF.  We find a power law IMF slope over the stellar
mass range \mrange\ of $\alpha = 1.2_{-0.5}^{+0.4}$ for Hercules and
$\alpha = 1.3\pm0.8$ for Leo IV, where $\alpha = 2.3$ for a
\citet{kroupa02a} IMF in this mass regime for the Milky Way.  Over our
mass range, the UFDs exhibit a shallow IMF deficient in low mass stars
relative to the Milky Way.

Comparing to other galaxies in which the IMF has been measured via
directly counting stars, we see a trend with galactic velocity
dispersion and metallicity.  The power law slope becomes increasingly
shallow (bottom-light) with decreasing galaxy velocity dispersion
and/or metallicity.  This trend is qualitatively consistent with
results in elliptical galaxies inferred via indirect methods
\citep{treu10a,vD11a, cappellari12a,dutton12a,conroy12a}.   The combined data provide
clear evidence for IMF variations with galactic environment.  This has
significant implications for galaxy formation models which often
assume a Milky Way-like IMF.  A galaxy-dependent IMF affects estimates
of fundamental properties such as supernova feedback rates, stellar
masses and chemical abundances \citep[e.g.,][]{ferre13a}.  A critical
question for future studies is understanding the physical properties
which control the IMF as a function of galactic environment.

The UFD IMFs presented here cover a relatively limited range in
stellar mass.  The Milky Way exhibits a transition to a shallower IMF
slope just below our mass limits.  Testing whether a similar
transition exists in the UFDs at 0.5\Msun\ or lower requires deeper
{\it HST} imaging which can reasonably probe the IMF down to the
hydrogen burning limit in the closest UFDs.  Alternatively, an IMF
slope transition could have existed in the UFDs at higher stellar mass
than our observed range.  Since stars more massive than 0.77\Msun\
have long since evolved off the main sequence, testing this is
difficult, but signatures may be present in the detailed chemical
abundances of individual stars \citep{vargas13a, Tsujimoto11a}.  Both
methods are needed to fully characterize the shape of the IMF and thus
probe the physical conditions of star formation in these extremely old
and metal-poor systems.

The details of star formation and the resulting IMF have long been
predicted to depend on the physical properties of the stellar birth
cloud \citep[e.g.,][]{larson05a, hennebelle08a, krumholz11a,
  hopkins12a}.  It is remarkable that direct evidence for IMF
variations with galactic environment is only now being uncovered, and
underlines the utility of the UFD galaxies for extending the baseline
of physical conditions in which the IMF can be directly constrained.

\vskip 0.25cm

\acknowledgments

We thank John Bochanski, Charlie Conroy, Kevin Covey, Pieter van
Dokkum, Andreas K\"upper, Richard Larson, Stella Offner, Joel Primack,
Risa Wechsler and Beth Willman for productive conversations.  Support
for GO-12549 was provided by NASA through a grant from STScI, which is
operated by AURA, Inc., under NASA contract NAS 5-26555.  MG
acknowledges support from NSF grant AST-0908752 and the Alfred
P.~Sloan Foundation.  ENK acknowledges support from the Southern
California Center for Galaxy Evolution and partial support from NSF
grant AST-1009973.  R.R.M.~acknowledges support from CONICYT through
project BASAL PFB-06 and from the FONDECYT project N$^{\circ}1120013$.

\bibliographystyle{../tex_files/apj}
\bibliography{../tex_files/apj-jour,imf_bib.bib}

\begin{deluxetable}{llllll}
\tabletypesize{\tiny}
\tablecaption{Luminosity Function for Hercules and Leo IV}
\tablehead{
\colhead{F814W} & \colhead{Herc} & \colhead{Herc} & \colhead{Leo IV} &
\colhead{Leo IV}\\
\colhead{(STMAG)} & \colhead{$N_{*, \rm raw}$} & \colhead{$f_{*, \rm compl}$} & 
                      \colhead{$N_{*, \rm raw}$} & \colhead{$f_{*, \rm compl}$} \\
}
\startdata
   24.35 &    4 &     0.91 &    . &     . \\
   24.45 &    2 &     0.90 &    . &     . \\
   24.55 &    5 &     0.90 &    . &     . \\
   24.65 &    8 &     0.90 &    3 &    0.94 \\
   24.75 &   15 &     0.90 &   0 &    0.94 \\
   24.85 &    8 &     0.90 &    3 &    0.93 \\
   24.95 &    8 &     0.90 &    5 &    0.93 \\
   25.05 &   18 &     0.89 &    6 &    0.92 \\
   25.15 &   20 &     0.89 &    6 &    0.93 \\
   25.25 &   22 &     0.88 &    4 &    0.93 \\
   25.35 &   25 &     0.88 &   12 &    0.92 \\
   25.45 &   26 &     0.88 &    9 &    0.92 \\
   25.55 &   32 &     0.87 &   10 &    0.91 \\
   25.65 &   26 &     0.86 &    8 &    0.91 \\
   25.75 &   34 &     0.86 &   11 &    0.90 \\
   25.85 &   36 &     0.86 &   12 &    0.90 \\
   25.95 &   40 &     0.85 &   16 &    0.90 \\
   26.05 &   56 &     0.85 &   18 &    0.90 \\
   26.15 &   63 &     0.83 &   18 &    0.88 \\
   26.25 &   50 &     0.83 &   23 &    0.89 \\
   26.35 &   70 &     0.83 &   25 &    0.87 \\
   26.45 &   71 &     0.82 &   23 &    0.87 \\
   26.55 &   56 &     0.82 &   18 &    0.87 \\
   26.65 &   88 &     0.81 &   23 &    0.87 \\
   26.75 &   78 &     0.80 &   24 &    0.85 \\
   26.85 &   83 &     0.80 &   33 &    0.85 \\
   26.95 &   83 &     0.79 &   29 &    0.85 \\
   27.05 &   78 &     0.79 &   41 &    0.84 \\
   27.15 &   92 &     0.78 &   44 &    0.84 \\
   27.25 &   84 &     0.77 &   29 &    0.83 \\
   27.35 &   83 &     0.76 &   47 &    0.83 \\
   27.45 &   62 &     0.75 &   27 &    0.83 \\
   27.55 &   88 &     0.72 &   31 &    0.82 \\
   27.65 &   86 &     0.70 &   41 &    0.82 \\
   27.75 &   88 &     0.68 &   43 &    0.81 \\
   27.85 &   83 &     0.67 &   31 &    0.81 \\
   27.95 &   96 &     0.66 &   44 &    0.80 \\
   28.05 &   94 &     0.66 &   39 &    0.80 \\
   28.15 &   85 &     0.66 &   42 &    0.79 \\
   28.25 &   91 &     0.66 &   40 &    0.78 \\
   28.35 &  111 &     0.66 &   44 &    0.76 \\
   28.45 &  130 &     0.66 &   53 &    0.76 \\
   28.55 &  .      &    .        &   36 &   0.76  \\
   28.65 &  .      &    .        &   48  &   0.75 \\
   28.75 &  .      &    .        &   35  &   0.75 \\
\enddata
\tablecomments{The luminosity function and photometric completeness
  values.  We list values only for the magnitude range included in the
  IMF analysis.}
\end{deluxetable}

\end{document}